\documentclass[aps,prl,preprint,groupedaddress]{revtex4}

\usepackage{amssymb}
\usepackage{enumerate}
\usepackage{graphicx}
\newcommand{\be}{\begin{equation}}
\newcommand{\ee}{\end{equation}}
\newcommand{\beq}{\begin{eqnarray}}
\newcommand{\eeq}{\end{eqnarray}}
\newcommand{\ba}{\begin{align}}
\newcommand{\ea}{\end{align}}
\newcommand{\de}{\partial}
\newcommand{\bef}{\begin{figure}}
\newcommand{\eef}{\end{figure}}

\begin{document}

\title{How much information can one store in a non-equilibrium medium?}

\author{P. Coullet, C. Toniolo, and C. Tresser}

\thanks{P.C. is Professeur \`{a} l'Institut Universitaire de France}
\thanks{C.T. is partially supported by NSF}
\thanks{This paper is dedicated to Paul Clavin,  Lorenz Kramer,  and Yves Pomeau on the occasion of their sixtieth birthday.}

\address{P.Co. and C. To.: INLN, 1361 Route des Lucioles, 06560, Valbonne,
France.}
\address{C.Tr.: IBM, P.O. Box 218, Yorktown Heights, NY 10598, U.S.A.}

\date{\today}

\begin{abstract}
It has recently been emphasized again that the very existence of stationary stable localized structures with short range interactions might allow to store information in non-equilibrium media,  opening new perspectives on information storage.   We show how to use generalized topological entropies to measure aspects of the quantities of storable and non-storable information.
This leads us to introduce a measure of the long term stably storable information.
As a first example to illustrate these concepts, we revisit a mechanism for the appearance of  stationary stable localized structures that is related to the stabilization of fronts between structured and unstructured states (or between differently structured states).
\end{abstract}


\maketitle

\textbf{
It has recently been emphasized again that the very existence of stationary
stable localized structures with short range interactions might allow to
store information in non-equilibrium media, in a way that would be both
local and reversible. This has altogether opened new perspectives on
information storage. While the theory of information was put of firm basis
by Shannon as far back as 1948, the focus of this theory has been on models
for the information itself, for which the concept of information theoretic
entropy was proposed, and on transmission channels, for which the concept of
channel capacity, later extended to topological entropy, was put forward. We
show here that topological entropy, in its generalized form that
accommodates multidimensional times as in foliation and tiling theories,
characterizes nicely the amount of information that can be stored in
non-equilibrium media with local stability in time. By measuring also the entropy of
non-storable information, we get access to a measure of a fragility of the
system. This fragility can be understood as the lack of localization of
errors, hereby leading us to introduce a measure of the long term stably storable
information. As a first example to illustrate these concepts, we revisit a
mechanism for the appearance of  stationary stable localized structures that is
related to the stabilization of fronts between structured and unstructured
states (or between differently structured states). The theory is far less
complete for media of dimension greater that one, but the concepts and main
effects of dimension one seem to work well for two dimensional media for
reasons whose details elude us.
}

\section{Introduction}\label{sec:introduction}
Many extended systems excited far from equilibrium undergo a transition
from a spatially homogeneous state to a periodic pattern.
Classical examples are the discrete vibration modes of a string, locked at
its extrema, the rolls and hexagonal structures formed by convection in a
Rayleigh-B\'enard geometry, the Taylor-vortices in a Taylor-Couette system,
the Faraday's crispations on the surface of a parametrically forced disk of
liquid (or granular material), the Turing patterns developing in chemical or
biological systems where reaction and diffusion between different specimen
compete \cite{CrossHohen}.

With regard to data storage, it is clear that a homogeneous solution cannot
store information, and that a periodic pattern only contains information about
a wavenumber and the position of one vertex, none of which has much value, while a
localized structure with short range interaction would seem a much better candidate.
Nonetheless, it is exactly the ability of many physical systems to switch between
two (or more) different stable states which is commonly exploited for information
storage.
More specifically, as already explained in \cite{CouRieTre0}, for a quite general class
of systems we can use the bistability between an homogeneous and a cellular state,
which is distinguished from the usual bistability between two homogeneous states
\cite{KogaKuramoto}, to ensure the existence of stable localized structures and hence
to store information on a medium.

Static localized structures, seen as independent fragments of a structured
solution, have been studied in many different physical systems, ranging from
magnetic materials \cite{ODell}, to liquid crystals \cite{Pirkl} and to
chemical systems \cite{Lee}, \cite{Herschkowitz}. More recently, there has
been a new surge of interest in the context of optics, where localized and
controllable light spots have been considered as bits for information
encoding and processing (see for instance \cite{Foot2}, \cite{Foot1} and
\cite{Firth}).

In this paper we will address the problem of quantifying the quantity of
information that can be stored and is made available in a non-equilibrium
medium, and the associated problem of transition from poor to efficient
storage performances in contexts where bifurcations toward localized solution
is reasonably well understood.

The discussion to follow is quite general, but to illustrate the
mechanism we will here refer to a system described either by
variational or nonvariational  partial differential equations (PDEs) of the form:
\beq
\de_t u = -\frac{\de V}{\de u}- \nu \nabla^2 u - \nabla^4 u\,,
\label{swifthohen_eq}
\eeq
where $V= - \mu u^2/2 +u^4/4 - \eta u$, or
\beq
\de_t u &=& -\frac{\de V}{\de u} -v + D_u \nabla^2 u, \nonumber\\
\de_t v &=& -\gamma v + c u + D_v \nabla^2 v\,.
\label{reacdiff_eq}
\eeq
Equation (\ref{reacdiff_eq}) describes a chemical reaction, while (\ref{swifthohen_eq})
is a generalization of the Swift-Hohenberg model \cite{Tlidi}.

With solutions bi-asymptotic to an homogeneous solution, and a suitable
choice of the parameter values, we have written HELLO in Figure 1-a, using
Equation \ref{swifthohen_eq}, and equation \ref{reacdiff_eq} would have
worked as well. Even though the "HELLO"  in Figure 1-a corresponds to a
linearly stable solution, one expects that there be a drift in the presence
of any non-zero noise, and Figure 1-c confirms this concern, even if one
takes the best possible parameter values. Following the teaching of
\cite{CouRieTre0}, one can also use  solutions bi-asymptotic to a cellular
solution to store information. We thus have accordingly again written HELLO
in Figure 1-b, also using Equation \ref{swifthohen_eq} and the same optimal
parameter values used for Figure 1-a  (as discussed in  \cite{CouRieTre0};
see also below).  Figure 1-d, a snapshot after the same time used for Figure
1-c, indicates that now the drift due to noise is, for the least, far less of a problem. 
The noise used is uniform in $[-c,c]$ and the same non negligible value $c=0.2$ has been used in both cases. The effect of this noise on the solution is almost negligible (and in fact fully
negligible for all practical purposes if the noise remains within some
reasonable bounds) in the presence of a cellular substrate and one uses a sub-lattice to carry the information: the single cells are "locked" to their positions by the super-lattice that is not used to carry information (the sub-lattice constraint can be relaxed, as long as no large homogeneous patches are created). The same rigidity property evidently does not hold true on a homogeneous background. 
While we have used two dimensional simulation for the spectacular aspect and 
because that would be the dimension of most obvious practical relevance, the phenomenon displayed in Figure 1 exist in any dimension, and the remarks we have made one the difference of rigidities would apply as well. This qualitative argument furnishes an intuitive, but surely not rigorous
interpretation of the behavior. On the other hand, the formulation of a
precise $\delta$-dimensional theory, taking into account the effect of noise,
stands beyond the scope of what can be done now. This is not surprising given the complexity of the
problem even in the absence of noise. More precisely, some rigor can be obtained about the global theory of stationary solutions when $\delta =1$ but for greater $\delta$ or when time dependence is considered, there is no global theory available so far, even in the noiseless case.

\begin{figure}[htbp]
\centerline {\includegraphics[width=0.85\textwidth]{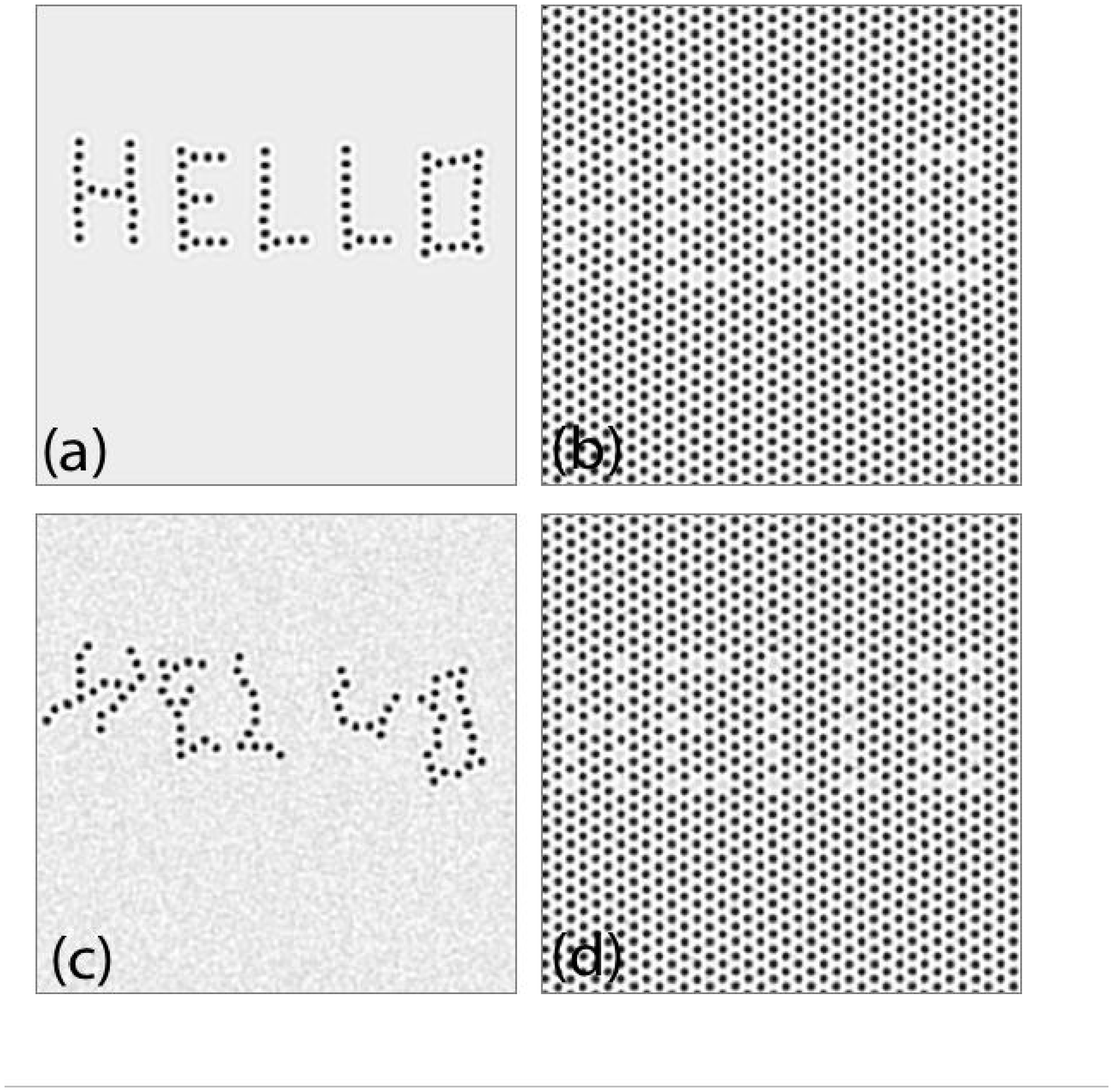}} \caption{
Writing the message "HELLO" using stable stationary localized structures made of bumps (to
the left) or of flats (to the right) on a medium of dimension two, using
Equation \ref{swifthohen_eq} to generate stable solutions. In (a) the message is coded with bumps on an homogeneous state; in (b)the message is coded with flats on a cellular substrate. In the lower panels (c) and (d) a nonzero noise is superposed: while a considerable drift
is observed in panel (c), the effect of noise is far less important in (d), and anyway smaller than the cells sizes. The parameters values of the simulations are: $\eta=-0.3, \mu=-0.4$ and
$\nu=-2$.  The noise in (c) and (d) is uniform in $[-0.2,0.2]$.} \label{Fig1}
\end{figure}

The bifurcation structure of stationary stable localized solutions near the
domain $\textbf{\textsl{P}}$ of existence of a stationary front between a cellular structure and
an homogeneous state was described in \cite{CouRieTre1} , \cite{CouRieTre2}
and \cite{CouRieTre0} in one-dimensional media: see Figure 2 which is
described in detail below.  As we move from the boundaries of the region of
existence of a localized solution to the interior of this region, and
further inside the region $\textbf{\textsl{P}}$, a variety of different
stable states is encountered depending on the parameter range. The quantity
of solutions that are available to the system becomes maximal in a special
parameter range $d$, already singled out in \cite{CouRieTre0}, were one finds
structures that were called \emph{fully decomposable}. For such systems with
parameters in $d$ (we say a \emph{decomposable system}), essentially any
cell of a lattice intrinsic to the system can be set ON or OFF in a very
recognizable way. Thus any information can be encoded with high density in the parameter range
$d$. There is a problem here with the lattice if big patches get turned off,
and it was suggested in  \cite{CouRieTre0} to use in fact a sublattice to
store information, and use the lattice also to recognize where one stands in
the medium. This difficulty will be reconsidered later, hidden behind the issue of
the \emph{discreteness quality of a system},  \emph{i.e.,} how well it can be
described as an assembly of bits. Cellular and homogeneous states clearly
stand at the other end of the spectrum from decomposable systems in terms of
efficiency in carrying information, and this despite the fact that all these
states can be reached with the same medium or with the same equations, by
just tuning the parameters. It was also documented in \cite{CouRieTre1} ,
\cite{CouRieTre2} and \cite{CouRieTre0} that similar phenomenology occurs in
two-dimensional media.

It has become clear, as we report here, that it would help a lot to
\emph{quantify} the information  that can be stored in a medium, somewhat in
analogy with the quantity of information that can be stored in a message as
described in Shannon's theory. Chaos theory and the language thereof have
greatly progressed folllowing the seminal paper of Claude Shannon
\cite{Shannon1948} (which can also be found  in \cite{ShannonWeaver1949} and
\cite{Shannon1993}) that put the theory of information on firm basis as far
back as 1948. While the focus of Shannon theory has been on models for the
information itself, for which the concept of information theoretic entropy
was proposed, and on transmission channels, for which the concept of channel
capacity, later extended to topological entropy, was put forward, we will be
able to build on these teaching, and in particular use Shannon's
quantitative approach for our problems of storage.

\section{Measuring spatial chaos}
We also learn from Shannon that it is convenient to begin with systems that
are discrete as they are somewhat simpler to analyze.  For a message, time
is discretized by defining a unit of time, and one then uses as elementary
information a symbol from a finite alphabet $\mathcal{A}$ (in most cases,
$\mathcal{A}=\{0,1\}$)  that is ascribed to each unit of time to form a
message. For spatial chaos, instead on the one-dimensional time, one has a
$\delta -$dimensional medium, with $\delta \geq 1$, so that  discretization
is more generally accomplished  by defining a  lattice.  We will assume that
in some parameter range, any cell of that lattice can similarly be ascribed
a symbol from a finite alphabet $\mathcal{A}$ (and we will start with,
$\mathcal{A}=\{0,1\}$ to fix the ideas), in such a way that a field can be
mimicked by a step function with constant value taken from this alphabet on
each cell.

The amount of spatial chaos, or amount of information that can be stored in the medium
with unit of space fixed by the lattice is the \emph{topological entropy} (in fact an obvious
 generalization of Shannon's \emph{channel capacity}), defined as:
\begin{equation}
C_{topological}(\lambda)=\lim_{n\to\infty}\frac{\log_2(N_\lambda(r))}{r^\delta}\,,
\end{equation}
where $N_\lambda(r)$ is the number of different patches that can be observed in a ball of
radius $r$ for the medium or equation with fixed parameter $\lambda$. Here $\lambda$ can be
 a scalar or more generally a vector, but to study transitions from zero to maximal entropy,
 we will stick to scalar parameters in the present paper. Our $C_{topological}(\lambda)$ is also
 the quantity used to measure the topological entropy of other spatial systems, such as tilings
 of the plane or more generally tilings of $\delta-$dimensional spaces.
 One should not be surprised to see topological entropy appear here since long after both
 concepts had been isolated (see \cite{Shannon1948} and  \cite{AdlerKonheimMcAndrew} ),
 it has finally been recognized that Shannon's channel capacity is the topological entropy of
 a subshift  of finite type. The historical filiation from \cite{Shannon1948}  to
 \cite{AdlerKonheimMcAndrew}  (see also \cite{Dinaburg} and \cite{Bowen}) is indeed through
 \cite{Kolmogorov} (see also \cite{Sinai}).  However, the Kolmogorov-Sinai entropy \cite{Kolmogorov},
 \cite{Sinai} was not so much inspired by the Shannon capacity, but rather by the Shannon entropy
 (see \cite{Shannon1948}, \cite{ShannonWeaver1949}, or \cite{Shannon1993}) for which Shannon
 was inspired by Tolman's dicussion of the entropy from statistical physics (see \cite{Tolman};
 the reverse influence of Shannon's theory on physics being later initiated by Brilloin \cite{Brilloin}).

Topological entropy thus seems to capture nicely the complexity of
information that can be stored with a stability that is local in space and
time. However, in the context of storage arises the question of dynamical
evolution of errors: if an error resulting from fluctuation bigger than the
range of stability occurs, one would like this error to not propagate. It is
fair to call \emph{forbidden state} a state that, under too strong but still
mild perturbation, degenerates in such a way that an unbounded amount of
data gets erased. The collection of such forbidden states can be measured by
an entropy $ C_{forbidden} $ defined just the same way as before, and we can
define the \emph{efficient entropy $C_{efficient}$} as the supremum of zero
and the difference between the measure of complexity and the size of the
forbidden set. Thus the actual quality of storage for a given parameter
$\lambda$ will be given by:
\beq
C_{efficient}(\lambda)=
\sup{[0,C_{topological}(\lambda) - C_{forbidden}(\lambda)}]\,.
\label{entropy_eq}
\eeq

In the upper portion of Figure 2 we have sketched, by making reference to
the system described by Equation \ref{swifthohen_eq}, the qualitative
behavior of the three entropies defined above with respect to the control
parameter $\lambda$. The topological entropy $C_{topological}$ is zero at
the exterior of region of existence of stationary stable localized solution. 
This  parameter range for $C_{topological}=0$ is slightly bigger and strictly contains the pinning region $\textbf{\textsl{P}}$. Now the topological entropy $C_{topological}$ is uniformly equal to one in the interior of $\textbf{\textsl{P}}$, where the number of observable
static states is maximal. But information storage is made more and more
effective only when one approaches the decomposability region
$d$, since along this approach the efficient entropy $C_{efficient}$ increases and the entropy
$C_{forbidden}$ of the forbidden sets decreases. Eventually, the noise fluctuations are no
more destructive inside the parameter range $d$ where the efficient entropy reaches its maximal
value, equal to one, since the forbidden entropy drops to zero.

\section{Stationary stable localized structures: the phenomenology}\label{sec:Phenomenology}

Having now means to measure chaos for stored information and stably stored
information, we can next describe examples which have motivated the relevant
entropy concepts; we refer the reader to \cite{CouRieTre0} for more details
(see also \cite{CouRieTre1}, and \cite{CouRieTre2})), but only the
phenomenology will be essential to our discussion, and not the mathematical
developments provided in those papers to justify and explain such
phenomenology.

\begin{figure}[htbp]
\centerline {\includegraphics[width=4in]{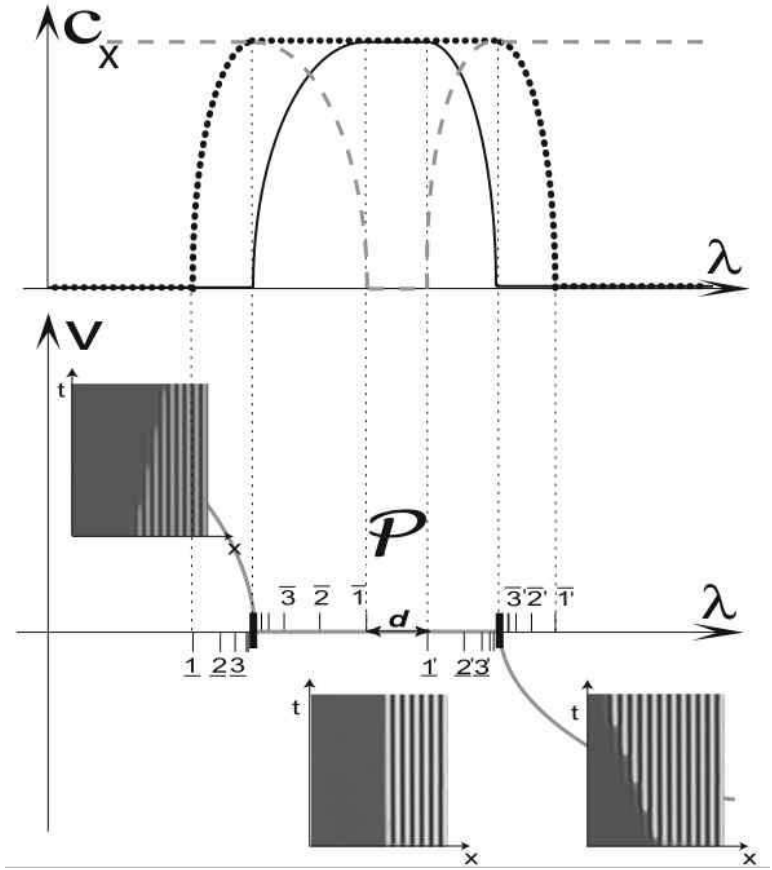}} \caption{
In the upper part of the figure, the three entropies $C_x$ used to measure spatial
chaos when the parameter $\lambda$ is varied: the thick dotted line is the topological
complexity $C_{topological}$, the dashed grey line represents the amount of forbidden
states $C_{forbidden}$, the solid thinner line is the efficient entropy $C_{efficient}$,
given by Equation \ref{entropy_eq}.
In the lower part of the figure, the bifurcation structure near the
pinning region $\textbf{\textsl{P}}$. The horizontal axis is the parameter
$\lambda$, the vertical axis the velocity $\textbf{v}$ of the front. Overbars
correspond to bubbles, underbars to flats ({\it e.g.}: $[\overline{1}, \overline{1}'$ is where a stationary stable localized strucrture with a single bump can be observed).
The sub-interval $d$ of $\textbf{\textsl{P}}$ is the region where the structure
is fully decomposable: this is where one should be for best data storage applications.
In the insets are three x-t diagrams  for $\lambda$ to the left, inside of, and to he right of $\textbf{\textsl{P}}$.
}
\label{Fig2}
\end{figure}

We will distinguish what we call  \emph{stable} and  \emph{unstable
landscapes}. As a default option, we will say \emph{landscape} for
\emph{stable landscape}. A \emph{landscape} is thus a stable solution up to
translation of the PDE, or more generally corresponds to a state that can
stably be observed on local fields in a non-equilibrium medium. Landscapes
are defined up to translation, so that a landscape corresponds to infinitely
many solutions: hence solutions are in one to one correspondence with
landscapes with a marked point, objects that we also call
\emph{configurations}. A \emph{state} is what a landscape (or a
configuration) locally looks like: thus landscapes can either exhibit the
same state all over, or a coexistence of states.

Consider a one dimensional medium which, depending on a control parameter
$\lambda$, can exhibit both a \emph{homogenous landscape} $\mathbf{H}$
(hence a \emph{stable homogenous state} all over the places) for low values
of $\lambda$, and a \emph{periodic landscape} $\mathbf{C}$ (hence a
\emph{stable cellular state} all over the places) for high values of
$\lambda$. We also assume that these states (whose precise description,
involving for instance the wavelength in the case of a cellular state, is
expected to depend on $\lambda$) coexist (meta-) stably in a single
parameter range $\textbf{\textsl{B}}$, that contains a \emph{pinning}
sub-range
$\textbf{\textsl{P}}=[_{\mathbf{C}\mapsto\mathbf{H}}\lambda,\lambda_{\mathbf{H}\mapsto\mathbf{C}}]$
where states that present a stable \emph{front} between the homogenous and
cellular state can be observed. For
$\lambda<_{\mathbf{C}\mapsto\mathbf{H}}\lambda$, the front is unstable with
the cellular state turning to homogeneous as the front moves (hence the
${\mathbf{C}\mapsto\mathbf{H}}$ \emph{left subscript} of
$_{\mathbf{C}\mapsto\mathbf{H}}\lambda$: see the leftmost inset in Figure 2,
where the vertical axis is time while the horizontal axis is along the
medium), while for $\lambda>\lambda_{\mathbf{H}\mapsto\mathbf{C}}$, the
front is \emph{unstable} with the homogeneous state turning to cellular as
the front moves (hence the ${\mathbf{H}\mapsto\mathbf{C}}$ \emph{right
subscript} of $\lambda_{\mathbf{C}\mapsto\mathbf{H}}$: see the rightmost
inset in Figure 2).

Mechanisms for the existence and robustness of such stable fronts were
discussed by Pomeau in \cite{Pomeau}, and revisited in \cite{CouRieTre1},
\cite{CouRieTre2}  using the concept of \emph{reversible systems} which was
isolated (with no view on application to PDE's) by Devaney \cite{Devaney1} in
the context of maps and ODE's. It was also explained in \cite{CouRieTre1},
\cite{CouRieTre2}, and \cite{CouRieTre0} that for some mechanism at least
that allows the phenomenology described so far, these phenomena come
accompanied with a complicated intertwining of parameter intervals
$I_\alpha$, that carry different possible stable states $\mathbf{S}_\alpha$
so that in the intersection of $I_{\alpha _1}$ and $I_{\alpha _2}$, both of
the states $\mathbf{S}_{\alpha _1}$ and $\mathbf{S}_{\alpha _2}$ can be
stably realized.

\medskip
We will need some vocabulary and notation to discuss respective positions of
intervals and phenomena, as this will be essential to understand the
parameter dependent phenomenology of stable stationary solutions. Thus, if
two intervals intersect in such a way that none contains the other one and
so that they share no extreme points, we say that one is \emph{kicked} (or
more precisely \emph{kicked to the right or two the left}) of the other one
(which is then of course kicked respectively to the left or right of the
first one). For instance the interval $[2,5]$ is kicked to the right of
$[1,3]$, and we notice that these two intervals have different length, which
is possible but not necessary when one interval is obtained from another one
by a \emph{kick}.

We are going to consider two types of localized states:

 - the \emph{$n$-bumps states} which correspond to a succession of exactly $n$ bumps with flats
 on both sides (after some oscillatory damping tail),

 - the \emph{$m$-flats states} which correspond to a succession of at least $m$ and less than $m+1$
 flats of the unit length that is the size of the bumps, with bumps on both sides
 (after some oscillatory expansion tail).

\noindent One can of course code the patterns that will be
possible by associating loosely a 1 to a bump and a 0 to a flat,
and more precisely:

 - a $1^n$ to an $n$-bumps state read to the right,

 - a $^{-n}1$ to an $n$-bumps state read to the left,

 - a $0^m$ to an $m$-flats state read to the right,

 - a $^{-m}0$ to an $m$-flats state read to the left,

\noindent where ``to the left" and ``to the right" means to the left and to the right of a reference point,
represented by the next symbol respectively to the right and left of the superscripted symbol. Using minus
signs allows a local reading: it allows to recognize if a superscript is attached to the symbol to
its left or to its right without having to see the whole word, assuming even such view is at all possible.

Then a \emph{symbolic configuration} on the real line $\mathbb{R}$, which is
everywhere made of bumps and flats can be represented by a bi-infinite word
in the \emph{alphabet} $\mathcal{A}_2=\{0,1\}$. For instance, the
homogeneous state $\mathbf{H}$ is represented by $^{-\infty}0^\infty$ and
the cellular state $\mathbf{C}$ by $^{-\infty}1^\infty$ while stable fronts
appear as $^{-\infty}01^\infty$ and $^{-\infty}10^\infty$. Local segments of
the medium can similarly be coded by finite words in $\mathcal{A}_2$ and in
most cases, one can as well dispense of negative exponents as one can always
read to the right when dealing with finite words (negative exponents may
however reveal handy when one intends to take some limits).

In Figure 2 the horizontal axis is the parameter $\lambda$ while the
vertical axis is the mean velocity $\textbf{\textsl{v}}$ of the front, which
is thus zero on the interval
$\textbf{\textsl{P}}=[\lambda_{\mathbf{C}\mathbf{H}},\lambda_{\mathbf{H}\mathbf{C}}]$
where there are stable static fronts between the states $\mathbf{H}$ and
$\mathbf{C}$.  The numbers with overbars correspond to bubbles, while the
numbers with underbars correspond to flats. More precisely:

\noindent - $I_{^{-\infty}01^n0^\infty}=[\overline{n}, \overline{n}']$ is where a stable localized solution,
symbolized as $^{-\infty}01^n0^\infty$ (or $^{-\infty}0^{-n}10^\infty$), with exactly $n$ consecutive
bubbles can be observed,

\noindent - $I_{^{-\infty}10^m1^\infty}=[\underline{m},\underline{m}']$ is
where a stable localized solution, symbolized as $^{-\infty}10^m1^\infty$
(or $^{-\infty}1^{-m}01^\infty$), with exactly $m$ consecutive flats can be
observed.

\noindent Furthermore:

-  for any two positive integers  $p$ and $q$ with $q>p$, $[\overline{q}, \overline{q}']$
is kicked to the left of $[\overline{p}, \overline{p}']$;

-  for any two positive integers  $p$ and $q$ with $q>p$, $[\underline{q}, \underline{q}']$
is kicked to the right of $[\underline{p}, \underline{p}']$;

-  for any four positive integers  $n$, $m$, $p$, and $q$ $[\overline{m}, \overline{n}']$
is kicked to the right of $[\underline{p}, \underline{q}']$.

\noindent At last, the sequences $\{\overline{n}\}$ and $\{\underline{m}\}$ accumulate on
$_{\mathbf{C}\mapsto\mathbf{H}}\lambda$ respectively from the right and from the left, while the
sequences $\{\overline{n}'\}$ and $\{\underline{m}'\}$ accumulate on
$\lambda_{\mathbf{H}\mapsto\mathbf{C}}$ respectively from the right and from the left.

As discussed in \cite{CouRieTre0}, the interval $[\overline{1}, \underline{1}']$, which is marked
$d$ on Figure 2, is the parameter region where the structure is fully \emph{decomposable} in the
sense that packets of bumps or holes can essentially be created at will (when and where needed).
It was argued in \cite{CouRieTre0} that this parameter range is where information storage and
retrieval would be most efficient; otherwise speaking, $d$ is the parameter range that one should
pick for data storage applications. We will see here that the quality of storage, as well as the
spatial chaos, can be measured by the entropies that we have defined, giving a more precise viewpoint
on what was proposed in \cite{CouRieTre0}.

\bigskip
\section{Transition to chaos: landscape complexity}
The complexity of the observable landscapes on $\delta$-dimensional media  increases when we
move in the parameter space from the cellular to the chaotic state, and decreases back to
zero when we reach the homogeneous state. Even though the phenomenology in the two dimensional
case looks similar to that which one observes in dimension one, the mathematical analysis of
the mechanisms behind the phenomena has so far only been developed in the latter case.
For this reason we describe the transition to chaos in spatial dimension $\delta =1$, using the notations we have developed to segment the $\lambda$ parameter space.

As we drive $\lambda$ inside the Pomeau region, $C_{topological}(\lambda)$ undergoes a continuous
transition: it starts from $0$ far enough at the exterior of the pinning region and
it becomes uniformly equal to $1=log_2 2$ over all of $\textbf{\textsl{P}}$.  More precisely:

The set of landscapes in interval $(\overline{2'}, \overline{1'}]$ is the set of orbits of
the sub shift of finite type generated by $\{10^{m'_1},0\}$ for some $m_1$ that decreases as
one gets (to the left in Figure 2) toward the pinning region.

When crossing $\overline{2'}$ while moving the parameter towards $\textbf{\textsl{P}}$, words
containing $11$ begin to be permissible, and the minimal number $m'_2$ of consecutive zero's
after a $11$ block decreases as one gets toward the pinning region.

With no orbit loosing allowability till one reaches the boundary of the pinning region,
when crossing $\overline{n'}$, words containing $1^{n'}$ begin to be permissible, and the
minimal number $m'_{n'}$ of consecutive zero's after a $1^{n'}$ block decreases as one gets
toward the pinning region.

A similar description works when moving in the other direction on the other
side of the pinning region, with the roles of 0's and 1's exchanged , and
$\underline{n}$ replacing $\overline{n'}$. For instance, the set of
landscapes in interval $[\underline{1}, \underline{2})$ is the set of orbits
of the subshift of finite type generated by $\{01^{m_1},1\}$, for some $m_1$
that decreases as one gets  (to the right in Figure 2) toward the region
$\textbf{\textsl{P}}$.

The entropy thus increases as $\lambda$ gets toward $\textbf{\textsl{P}}$, and is uniformly 1
in $\textbf{\textsl{P}}$ as was told.

The description given before of the intervals of the form $[\underline p,
\underline p']$ or $[\overline q, \overline q']$ allow then to see that,
when considering $\lambda\in\textbf{\textsl{P}}$, the entropy of forbidden
words is zero exactly in $[\overline 1, \underline1']$. This entropy
increases on both sides of $[\overline 1, \underline1']$, when moving
towards the bounds of $\textbf{\textsl{P}}$, reaching 1 precisely at the
extreme points of $\textbf{\textsl{P}}$.

Hence, not all the entropy measured in the Pomeau region can effectively be
used to handle information. Even inside the pinning region there are still
landscapes that are not accessible for information applications, the optimal
parameter set being  the decomposability region $d=[\overline 1,
\underline1']$, which is precisely the locus of parameter values where
$C_{efficient}(\lambda)= 1$.

At this point, we should get back to the observed better quality of using
coding by 0's (for flats) on a 1's (for bumps) basis over the other way
round (see Figure 1 and the accompanying comments). On the other hand, the story is totally symmetric between 0's and 1's at the symbolic level under the proper change of parameter. More precisely,
for any phenomenology with 0's and 1's seen for some $\lambda$, there should
be a symmetric phenomenology with 1's and 0's that one should expect to see
at some other value of the parameter value $\lambda '$. What seems to be a
contradiction is due to the fact that all the analysis that we have
presented so far has been made under the hypothesis that there was formation
of a lattice, {\it i.e.,} under the hypothesis that coding is well defined. For a lattice to lead to a symmetry between 0's and 1's, the
lattice has to have cells that behave as stably when they are marked 0 or 1.
This symmetry happens to be much more the case, with an effective quantization of the
lengths of the cells, when there is an overall lattice of bumps with a
sub-lattice devoted to carry the information as was proposed in
\cite{CouRieTre0}. Otherwise speaking, the more one gets closer to the
parameter range $d$, the more the assumption that there is a good coding
with 0's and 1's is true.  The theory that we have developed takes in effect all its sense
in $d$ with information-carrying sub-lattices on a structured substrate,
where the theory is most important for applications, and where the
limitation on the effective entropy does not come from forbidden
configurations, but from the fact that one use only a sub-lattice for
coding. For instance, in $d$ when $\delta =1$, if one uses only one site out
of two for coding, the effective entropy available for information is
$\frac{1}{2}$ and in general, in any dimension, this effective entropy
decreases proportionally to the density of the information carrying
sub-lattice.

\bigskip
\section{Conclusion}
In this paper we have made the first step to the formulation of a coherent theory of classical
information storage. At this point, we only have the rudiments of a stability theory for the stationary
case when the storage medium has dimension one. For that, we needed some measurable quantities which we expect to be useful far beyond the scope of where we have theoretical control.
It would be nice to have an extension of the theory of information storage to more general cases,
including higher dimension and  spatio-temporal patterns (so that the dynamics may help for instance
in aspects of information processing), perhaps also to quantum information.

\bigskip
\noindent \textbf{Acknowledgments.} This work is partially
supported by the European Community grants EP 28235 - PIANOS and
FMRX-CT96-0010, and by NSF Grant DMS-0073069. The simulations were done with the XDim Interactive Simulation package developed by Pierre Coullet and Marc Monticelli. C. Riera is the author of the spectral code used for the resolution of equation 1.

\bigskip
\end{document}